\documentclass[a4paper,11pt]{article}
\usepackage{pos}
\usepackage{amsfonts}
\usepackage{amsmath}
\usepackage{braket}

\newcommand{\llangle}{\Big\langle \!\! \Big\langle}
\newcommand{\rrangle}{\Big\rangle \!\! \Big\rangle}
\newcommand{\xx}{\underline{x}}

\newcommand{\kk}{\underline{k}}

\newcommand{\qq}{\underline{p}}
\newcommand{\dd}{\text{d}}

\title{Constraining Proton Spin at Small $x$ with Valence Quark Model}

\author*[a]{Yossathorn Tawabutr}
\author[b]{Daniel Adamiak}
\author[c,d]{Heikki Mäntysaari}

\affiliation[a]{Center of Excellence in High Energy Physics, Faculty of Science, Chulalongkorn University,\\
  254 Phaya Thai Rd, Wang Mai, Pathum Wan, Bangkok 10330, Thailand}

\affiliation[b]{Michigan State University,\\
East Lansing, Michigan 48824, USA}

\affiliation[c]{Department of Physics, University of Jyväskylä,\\
P.O. Box 35, 40014 University of Jyväskylä,
Finland}

\affiliation[d]{Helsinki Institute of Physics,\\
P.O. Box 64, 00014 University of Helsinki, Finland}

\emailAdd{yossathorn.t@chula.ac.th}

\abstract{
Recently, a global analysis has been performed~\cite{Adamiak:2025mdy} for longitudinal spin asymmetries in polarized deep inelastic scattering (DIS) processes, combining the polarized valence quark model~\cite{Dumitru:2024pcv} with the large-$N_c$\&$N_f$ helicity evolution equations~\cite{Cougoulic:2022gbk}. With a three-fold reduction in the number of free parameters, the valence quark model results in a six-fold reduction of uncertainty for the resulting prediction of total parton spin inside the proton, when compared to previous analysis~\cite{Adamiak:2023yhz} performed with traditional moderate-$x$ model inspired by Born approximation. The fit based on valence quark model predicts a positive gluon hPDF and negative $g_1$ structure function at small $x$. However, the resulting increase in the $\chi^2$-statistic of the fit warrants further examination into the physical uncertainty of the valence quark model, which is a work in progress. We also summarize other possibilities of future research directions towards the end of this article.
}

\FullConference{The 33rd International Workshop on Deep Inelastic Scattering and Related Subjects (DIS2026)\\
4 - 8 May 2026\\
Bologna, Italy\\}

\begin{document}
\maketitle

\section{Motivations and Background}
\label{sec:intro}

The \emph{proton spin puzzle} has been an intriguing problem in strong interaction physics, starting from the measurement~\cite{Ashman:1987hv,Ashman:1989ig} by the European Muon Collaboration (EMC) that found a significant mismatch between the proton spin of $\frac{1}{2}$ and the total spin carried by the quarks inside. This generalizes our understanding of proton spin to the \emph{spin sum rules,}~\cite{Jaffe:1989jz,Bashinsky:1998if,Leader:2013jra}
\begin{align}
\label{JM_sum_rule}
    \frac{1}{2} = S_q + S_g + L_q + L_g\,,
\end{align}
where $S_q$ ($S_g$) is the total spin of the quarks (gluons) inside the proton and $L_q$ ($L_g$) is the total orbital angular momentum (OAM) of the quarks (gluons). This article focuses on the first two terms of Eq.~\eqref{JM_sum_rule} in the helicity basis, for which the respective terms can be written as
\begin{align}
\label{Sq_Sg}
    S_q(Q^2) &= \frac{1}{2}\int\limits_0^1\dd{x}\,\Delta\Sigma(x,Q^2)\;\;\;\;\;\text{and}\;\;\;\;\;S_g(Q^2) = \int\limits_0^1 \dd x\,\Delta G(x,Q^2)\,.
\end{align}
Here, $\Delta\Sigma$ and $\Delta G$ are respectively the (flavor-singlet) quark and gluon helicity distribution functions (hPDFs). The kinematic variables include the Bjorken $x$ and the photon virtuality $Q^2$. One of the challenging parts of Eq.~\eqref{Sq_Sg} lies in the lower limit of each integral over $x$, which goes all the way to $x=0$. Partons at such small values of Bjorken $x$ can only be probed via deep inelastic scattering (DIS) at asymptotically large center-of-mass (CM) energy, making it difficult to obtain a complete knowledge of parton helicity entirely from experimental measurements.

An alternative route in obtaining information about the hPDFs at small Bjorken $x$ lies within the color glass condensate (CGC) framework~\cite{Gelis:2010nm,Kovchegov:2012mbw}, which allows for distribution functions at small $x$ to be written in terms of the amplitude for a color-singlet quark-antiquark dipole to interact with the proton target at high energy. This quantity is called the \emph{dipole amplitude.} Via various methods -- including the background field method~\cite{Balitsky:2008zz} and the light-cone operator treatment (LCOT)~\cite{Cougoulic:2022gbk,Adamiak:2024khm} -- the evolution equation in rapidity can be derived for the dipole amplitude, allowing for distribution functions at small $x$ to be determined from their profiles at moderate $x$~\cite{Kovchegov:2012mbw}. The latter can be extracted from experimental measurements, c.f.~\cite{Albacete:2010sy,Beuf:2020dxl}.

For the hPDFs, the matching process between the moderate-$x$ profile and experimental measurement requires a fit to a relatively limited amount of data. As seen in the results of recent analyses~\cite{Adamiak:2021ppq,Adamiak:2023yhz,JAMCollaborationSmall-xAnalysisGroup:2025tfa}, the helicity evolution equation~\cite{Kovchegov:2015pbl,Kovchegov:2016zex,Kovchegov:2018znm,Kovchegov:2021lvz,Cougoulic:2022gbk} is capable of describing the small-$x$ polarized DIS and semi-inclusive DIS (SIDIS) data, but the uncertainties of model parameters for moderate-$x$ initial conditions of the evolution remain significant. This is a symptom of an overfitting problem, which can be resolved with either more experimental data or a more deterministic model. Although the former will improve drastically with the upcoming electron-ion collider (EIC)~\cite{Accardi:2012qut,AbdulKhalek:2021gbh,Abir:2023fpo}, the latter can be achieved today by introducing physical features of proton into the model.

An attempt in this direction comes in the form of the valence quark model, in which the proton is approximated to faithfully contain the constituent two up quarks and one down quark. First formulated in~\cite{Dumitru:2018vpr,Dumitru:2020fdh,Dumitru:2020gla}, the model expresses dipole amplitude in terms of the valence-quark wave function of the proton in the light-cone coordinates, with perturbative gluon emission included. The last part bridges the gap between the strict valence-quark regime of $x\sim 0.3$ and the ``moderate-$x$'' regime relevant for the initial conditions of small-$x$ evolution equations, which corresponds to $x\sim 0.01$ for unpolarized~\cite{Albacete:2010sy} and $x\sim 0.1$ for helicity evolution~\cite{Adamiak:2021ppq}.\footnote{This is partly a consequence of the fact that unpolarized BK/JIMWLK evolution resums $\alpha_s\ln(1/x)$, while the helicity evolution resums $\alpha_s\ln^2(1/x)$.} Subsequently, the valence quark model has been extended to hPDFs~\cite{Dumitru:2024pcv}, whose initial-condition regime of $x\sim 0.1$ is much closer to $x\sim 0.3$. The physical input from the valence quarks cuts down the number of free parameters to 1/3 the original number for the moderate-$x$ model of hPDFs, providing a promising component for the new global analysis~\cite{Adamiak:2025mdy} of hPDFs at small $x$, which is the main topic of this article.

In subsequent sections, necessary ingredients and features of small-$x$ helicity evolution are summarized in Section~\ref{sec:helicity_evol}. Then, Section~\ref{sec:VQ} outlines the valence quark model calculation for hPDFs~\cite{Dumitru:2024pcv}, whose results have been employed in a new global analysis~\cite{Adamiak:2025mdy} that is presented in Section~\ref{sec:globalfit}. Finally, we conclude and provide future directions of this development in Section~\ref{sec:conclusion}. 

Throughout the article, we employ the Einstein's summation convention with mostly minus Minkowski metric. In addition, the light-cone coordinates are employed with $v^{\pm}=\frac{1}{\sqrt{2}}(v^0\pm v^3)$, together with the notation, $\underline{v}=(v^1,v^2)$ for transverse vectors. Transverse separation vectors are sometimes labeled as $\xx_{ij} = \xx_i-\xx_j$, with $x_{ij} = |\xx_{ij}|$.

\section{Landscape of Parton Helicity at Small $x$}
\label{sec:helicity_evol}

Within the CGC framework, quark and gluon hPDFs relate to the contributions to dipole amplitude that are generally suppressed by a power of center-of-mass energy, $s$~\cite{Kovchegov:2015pbl,Kovchegov:2018znm,Cougoulic:2022gbk}. This is called the \emph{sub-eikonal} order. All relevant contributions to hPDFs come in the form of insertions of one energy-suppressed vertex to the leading-$s$ -- \emph{eikonal} -- dipole amplitude. Such vertices come in the form of (i) type-1 quark-exchange vertices, (ii) type-1 gluon-exchange vertex and (iii) type-2 gluon-exchange vertex.\footnote{More recently, quark-to-gluon and gluon-to-quark vertices are found to provide significant contribution, albeit at N$^3$LO in DGLAP power counting~\cite{Borden:2024bxa}.} If we represent as fundamental Wilson line $V_{\underline{x}}$ the high-energy interaction between a quark at transverse position $\underline{x}$ and the target, as per the usual CGC prescription, then the three types of sub-eikonal quark-target interactions can be represented by $V_{\underline{x}}^{\text{q}[1]}$, $V_{\underline{x}}^{\text{G}[1]}$ and $V_{\underline{x}}^{i\,\text{G}[2]}$, respectively~\cite{Cougoulic:2022gbk}. The three types of sub-eikonal Wilson line define the following sub-eikonal \emph{polarized dipole amplitudes:} 
\begin{subequations}\label{pol_dip_amp}
\begin{align}
    &Q^q_f(x^2_{10},zs) = \int \dd^2\left(\frac{\xx_1+\xx_0}{2}\right) \frac{1}{2N_c}\, \text{Re} \llangle \mathcal{T}\,\text{tr}\left[V_{\xx_0}V_{\xx_1}^{\text{q}[1]\dagger}\right] + \mathcal{T}\,\text{tr}\left[V_{\xx_1}^{\text{q}[1]}V_{\xx_0}^{\dagger}\right]\rrangle (zs) \,  , \label{Qq} \\
    &Q^G(x^2_{10},zs) = \int \dd^2\left(\frac{\xx_1+\xx_0}{2}\right) \frac{1}{2N_c}\,\text{Re} \llangle \mathcal{T}\,\text{tr}\left[V_{\xx_0}V_{\xx_1}^{\text{G}[1]\dagger}\right] + \mathcal{T}\,\text{tr}\left[V_{\xx_1}^{\text{G}[1]}V_{\xx_0}^{\dagger}\right]\rrangle (zs) \,  , \label{QG} \\
    &G_2(x^2_{10},zs) = \frac{\epsilon^{ij}\xx_{10}^j}{x^2_{10}} \int \dd^2\left(\frac{\xx_1+\xx_0}{2}\right) \frac{1}{2N_c}\, \llangle \text{tr}\left[V_{\xx_0}^{\dagger}V_{\xx_1}^{i\,\text{G}[2]}\right] + \text{tr}\left[V_{\xx_1}^{i\,\text{G}[2]\dagger}V_{\xx_0}\right]\rrangle (zs) \,  , \label{G2}
\end{align}
\end{subequations}
where $x_{10}$ is the dipole size and $zs$ is the longitudinal momentum fraction of the polarized (anti)quark in the dipole. Here, $\langle \! \langle \cdots \rangle \! \rangle (zs) = zs\left\langle\cdots\right\rangle$, while $\left\langle\cdots\right\rangle = \sum_SS\frac{\bra{P,S}\cdots\ket{P,S}}{\braket{P,S|P,S}}$ represents the \emph{helicity-dependent CGC-average} over the state of the proton target. Note that only $Q^q_f$ is flavor-dependent. In addition, the type-1 amplitudes also have non-trivial flavor non-singlet counterparts:
\begin{subequations}\label{pol_dip_amp_NS}
\begin{align}
    &Q^{\text{NS},q}_f(x^2_{10},zs) = \int \dd^2\left(\frac{\xx_1+\xx_0}{2}\right) \frac{1}{2N_c}\, \text{Re} \llangle \mathcal{T}\,\text{tr}\left[V_{\xx_0}V_{\xx_1}^{\text{q}[1]\dagger}\right] - \mathcal{T}\,\text{tr}\left[V_{\xx_1}^{\text{q}[1]}V_{\xx_0}^{\dagger}\right]\rrangle (zs) \,  , \label{Qq_NS} \\
    &Q^{\text{NS},G}(x^2_{10},zs) = \int \dd^2\left(\frac{\xx_1+\xx_0}{2}\right) \frac{1}{2N_c}\,\text{Re} \llangle \mathcal{T}\,\text{tr}\left[V_{\xx_0}V_{\xx_1}^{\text{G}[1]\dagger}\right] - \mathcal{T}\,\text{tr}\left[V_{\xx_1}^{\text{G}[1]}V_{\xx_0}^{\dagger}\right]\rrangle (zs) \,  . \label{QG_NS}
\end{align}
\end{subequations}
Altogether, these amplitudes combine to yield the quark hPDF of
\begin{subequations}\label{quark_hPDF}
\begin{align}
    &\Delta\Sigma(x,Q^2) = \sum_f\left[\Delta f(x,Q^2)+\Delta\Bar{f}(x,Q^2)\right] = \int\limits^{Q^2}\dd k^2_{\perp}\, g_{1L}^{\text{S}}(x,\kk) \label{singlet_hPDF} \\
    &= -\frac{N_c}{2\pi^3}\sum_f \int\limits_{\Lambda^2/s}^1\frac{\dd z}{z} \int\limits_{1/zs}^{\min\{1/zQ^2,1/\Lambda^2\}}\frac{\dd x^2_{10}}{x^2_{10}} \left[Q_f^q(x^2_{10},zs)+Q^G(x^2_{10},zs)+2G_2(x^2_{10},zs)\right] , \notag \\
    &\Delta q^-_f(x,Q^2) = \Delta f(x,Q^2)-\Delta\Bar{f}(x,Q^2) = \int\limits^{Q^2}\dd k^2_{\perp}\, g_{1L}^{\text{NS}}(x,\kk) \label{nonsinglet_hPDF} \\
    &= \frac{N_c}{2\pi^3}\sum_f \int\limits_{\Lambda^2/s}^1\frac{\dd z}{z} \int\limits_{1/zs}^{\min\{1/zQ^2,1/\Lambda^2\}}\frac{\dd x^2_{10}}{x^2_{10}} \left[Q_f^{\text{NS},q}(x^2_{10},zs)+Q^{\text{NS},G}(x^2_{10},zs) \right] , \notag
\end{align}
\end{subequations}
for the flavor singlet and non-singlet cases, respectively. Here, $\Lambda$ is the usual confinement scale of strong interaction. For the gluon hPDF, we have
\begin{align}
\label{gluon_hPDF}
    \Delta G(x,Q^2) = \frac{2N_c}{\alpha_s\pi^2} \, G_2(x^2_{10},Q^2/s)\, .
\end{align}
In the Veneziano limit of large $N_c$ and $N_f$~\cite{Veneziano:1976wm}, the polarized dipole amplitudes~\eqref{pol_dip_amp} form a closed system of integro-differential equations~\cite{Cougoulic:2022gbk} together with the polarized adjoint amplitude:
\begin{align}
\label{G_tilde}
    \widetilde{G}(x^2_{10},zs) &= N_fQ^G(x^2_{10},zs) \\
    &+ \sum_f\int \dd^2\left(\frac{\xx_1+\xx_0}{2}\right) \frac{1}{2N_c}\,\text{Re} \llangle \mathcal{T}\,\text{tr}\left[V_{\xx_0}W_{\xx_1}^{\text{q}[1]\dagger}\right] + \mathcal{T}\,\text{tr}\left[W_{\xx_1}^{\text{q}[1]}V_{\xx_0}^{\dagger}\right]\rrangle (zs) \,  , \nonumber
\end{align}
where $W_{\xx_1}^{\text{q}[1]}$ represents the sub-eikonal quark-exchange vertices on the gluon interacting with the target, corresponding to an adjoint Wilson line. The system of equations -- the small-$x$ helicity evolution equations -- allow us to access the polarized dipole amplitudes at small $z$'s, which are necessary to calculate the hPDFs at small $x$'s. 

As is the case for any differential equation, the small-$x$ helicity evolution requires an \emph{initial condition} in order to completely determine the solution. Because the evolution resums $\alpha_s\ln^2(1/x)$ per iteration, the initial conditions come in the form of 
\begin{align}
\label{F_ic}
    F^{(0)}(x^2_{10},zs) = a_F\ln\frac{zs}{\Lambda^2} + b_F\ln\frac{1}{x^2_{10}\Lambda^2} + c_F \,, 
\end{align}
for $F\in\{Q_f,G_2,Q^{\text{NS}}_f,\widetilde{G}\}$ with $Q_f=Q_f^q+Q^G$ and $Q^{\text{NS}}_f=Q^{\text{NS},q}_f+Q^{\text{NS},G}$. Traditionally, parameters -- $a_F$, $b_F$ and $c_F$ -- are taken to be free parameters of the global fits~\cite{Adamiak:2021ppq,Adamiak:2023yhz,JAMCollaborationSmall-xAnalysisGroup:2025tfa}. With three light quark flavors, $f\in\{u,d,s\}$, the number of free parameters rack up to 24 if polarized SIDIS data are to be included~\cite{Adamiak:2023yhz,JAMCollaborationSmall-xAnalysisGroup:2025tfa}, potentially leading to overfitting issues discussed in Section~\ref{sec:intro}. In this context, he polarized valence quark model proposed in~\cite{Dumitru:2024pcv} is capable to completely fix 16 of the 24 parameters via physical picture of a proton. The construction and utilization of the model will be discussed in Sections~\ref{sec:VQ} and \ref{sec:globalfit}.




\section{Valence Quark Model for Proton at Moderate $x$}
\label{sec:VQ}

As evident from Eqs.~\eqref{pol_dip_amp} and \eqref{G_tilde}, polarized dipole amplitudes involve helicity-dependent CGC-averaging of sub-eikonal Wilson lines, which essentially contain products of quark and gluon field operators. The latter can be written as color currents, which via strong-interaction equation of motion yields products of quark field operators. Ultimately, one needs to evaluate products of quark ladder operators sandwiched between the proton state, which within the valence quark model can be written as~\cite{Dumitru:2024pcv,Dumitru:2018vpr,Dumitru:2020fdh,Dumitru:2020gla}
\begin{align}
\label{ps_state}
    &\ket{P,S} = \frac{1}{\sqrt{6}}\int\frac{\dd x_1\,\dd x_2\,\dd x_3}{(4\pi)^3 \sqrt{x_1x_2x_3}} \, 4\pi\delta(1-x_1-x_2-x_3) \int\frac{\dd^2q_1\,\dd^2q_2\,\dd^2q_3}{(2\pi)^6}\,(2\pi)^2\delta^2(\qq_1+\qq_2+\qq_3) \notag \\
    &\;\;\;\;\times \sum_{f_1,f_2,f_3\in\{u,u,d\}} \sum_{\sigma_1,\sigma_2,\sigma_3} \Phi(x_1,x_2,x_3;\qq_1,\qq_2,\qq_3)\, \mathcal{S}_S(f_1,f_2,f_3;\sigma_1,\sigma_2,\sigma_3) \\
    &\;\;\;\;\times \sum_{i_1,i_2,i_3}\epsilon_{i_1i_2i_3} \ket{q(p_1,i_1,\sigma_1,f_1)} \otimes \ket{q(p_2,i_2,\sigma_2,f_2)} \otimes \ket{q(p_3,i_3,\sigma_3,f_3)} ,  \notag
\end{align}
where $\mathcal{S}_S(f_1,f_2,f_3;\sigma_1,\sigma_2,\sigma_3)$ is the totally symmetric spin-flavor state. Eq.~\eqref{ps_state} writes a state of proton with momentum $P$ and helicity $S$ as a totally antisymmetric (via color degree of freedom) state of two up and one down quarks. Here, $\Phi(x_1,x_2,x_3;\qq_1,\qq_2,\qq_3)$ is the kinematic part of the \emph{valence-quark wave function.} In light of the large center-of-mass energy regime in consideration, we employ the harmonic oscillator model by Schlumpf~\cite{Schlumpf:1992vq,Brodsky:1994fz}:
\begin{align}
\label{phi}
    \Phi(x_1,x_2,x_3;\qq_1,\qq_2,\qq_3) = \mathcal{N}\,\exp\left[-\frac{1}{2\beta^2}\sum_{i=1}^3\frac{q^2_{i\perp}+M^2}{x_i}\right],
\end{align}
where $\mathcal{N}$ is the normalization constant. The parameters are taken to be $\beta=0.55$ GeV and $M=0.26$ GeV, in order to be consistent with proton's radius and electromagnetic form factor~\cite{Schlumpf:1992vq,Brodsky:1994fz,Dumitru:2024pcv}.

With proton state specified by Eqs.~\eqref{ps_state} and \eqref{phi}, each polarized dipole amplitude can be calculated as a function of dipole size, $x_{10}$, and momentum fraction, $z$. The leading contributions from both quark- and gluon-exchange terms come at order $\mathcal{O}(\alpha_s^2)$. Both cases require an additional gluon exchange with the proton target due to an eikonal emission by the quark or antiquark in the dipole. Upon interpreting the ultraviolet cutoff as the center-of-mass energy of the (anti)quark-target interaction~\cite{Dumitru:2024pcv,Cougoulic:2022gbk,Kovchegov:2012mbw,Gelis:2010nm}, the valence quark model yields the coefficients of both transverse and energy logarithms -- $a_F$ and $b_F$ in Eq.~\eqref{F_ic} -- for each of the amplitudes. In particular, we have~\cite{Dumitru:2024pcv,Adamiak:2025mdy}
\begin{align}
\label{VQ_params}
    a_{Q_f} &= \frac{4\pi}{81}\alpha_s^2 \left[9 - 8\delta_{f,u} + 2\delta_{f,d} \right] \overline{x^{-1}} \, , \;\;\; b_{Q_f} = - \frac{2\pi}{81}\alpha_s^2 \left[18 + 4\delta_{f,u} - \delta_{f,d} \right] \overline{x^{-1}} \, , \\
    a_{Q_f^{\text{NS}}} &= - \frac{8\pi}{81}\alpha_s^2 \left[4\delta_{f,u} - \delta_{f,d} \right] \overline{x^{-1}} \, , \;\;\;\;\;\;\;\;b_{Q_f^{\text{NS}}} = - \frac{2\pi}{81}\alpha_s^2 \left[4\delta_{f,u} - \delta_{f,d} \right] \overline{x^{-1}} \, , \notag  \\
    a_{G_2} &= 0 \, , \;\;\;\;\;\;b_{G_2} = \frac{2\pi}{9} \alpha_s^2\, \overline{x^{-1}} \, , \;\;\;\;\;\;a_{\widetilde{G}} = \frac{23\pi}{18} \alpha_s^2\, \overline{x^{-1}}\;\;\;\;\;\text{and}\;\;\;\;\;b_{\widetilde{G}} = -\frac{11\pi}{9} \alpha_s^2\, \overline{x^{-1}} \, ,  \notag
\end{align}  
where $\overline{x^{-1}} = 3.64$ is the expectation value of the reciprocal of the longitudinal momentum fraction for a valence quark inside the proton, computed using the Schlumpf wave function \eqref{phi}. The numerical values in Eqs.~\eqref{VQ_params} are known up to the value of coupling constant, $\alpha_s$, which can be taken as a constant (``fixed coupling scheme'') or a function of a transverse position or momentum. The choice in this regard allows for customization of the fits in which the valence quark model is employed as initial conditions to the small-$x$ helicity evolution.




\section{Helicity Global Analysis with Valence Quark Model}
\label{sec:globalfit}

As discussed in Section~\ref{sec:VQ}, the valence quark model reduces the number of free parameters in the initial conditions from 24 to 8, with the prescription of $\alpha_s$ being the additional ``freedom'' for each fit. In Ref.~\cite{Adamiak:2025mdy}, the first global analysis based on the valence quark model~\cite{Dumitru:2024pcv} is performed within Jefferson Lab Angular Momentum (JAM) Monte Carlo Bayesian framework for the small-$x$ helicity evolution~\cite{Cougoulic:2022gbk}, comparing theoretical predictions to experimental measurements~\cite{E142:1996thl,E143:1998hbs,E154:1997xfa,E155:1999pwm,E155:2000qdr,EuropeanMuon:1989yki,SpinMuon:1997yns,SpinMuon:1998eqa,SpinMuon:1999udj,COMPASS:2009kiy,COMPASS:2010hwr,COMPASS:2010wkz,COMPASS:2015mhb,COMPASS:2016jwv,HERMES:1997hjr,HERMES:1999uyx,HERMES:2004zsh,HERMES:2006jyl} of longitudinal spin asymmetries in polarized DIS and SIDIS processes, which contain 226 data points in total. The small-$x$ evolution is taken to begin at $x=0.1$, owing to its double-logarithmic resummation~\cite{Adamiak:2021ppq,Adamiak:2023yhz,JAMCollaborationSmall-xAnalysisGroup:2025tfa}. Of the free parameters in Eq.~\eqref{F_ic}, the coefficients, $a_F$ and $b_F$, of the longitudinal and transverse logarithm are fixed exactly by Eq.~\eqref{VQ_params}, while the constant term, $c_F$, remains free. In addition, the strong coupling constant, $\alpha_s$, which appears with two powers in Eq.~\eqref{VQ_params}, is taken to follow the transverse dipole size:
\begin{align}
\label{as_VQ}
    \alpha_s(x_{10}) &= \frac{12\pi}{\left(33-2N_f\right) \ln\left[\frac{4C^2}{x^2_{10}\Lambda^2}\right]} \, ,
\end{align}
where the constant inside the logarithm, $C$, is introduced to account for the uncertainty of the exact value of $\Lambda$~\cite{Albacete:2010sy,Beuf:2020dxl}. The latter is taken to be $\Lambda=0.241$ GeV~\cite{Adamiak:2025mdy}. The running coupling prescription from Eq.~\eqref{as_VQ} is employed in the valence-quark initial conditions, in contrast to the prescription employed for $\alpha_s$ within the small-$x$ evolution equations, for which we employ the daughter-dipole prescription for consistency, following the choice of~\cite{Adamiak:2021ppq,Adamiak:2023yhz}. Specifically, the daughter-dipole prescription follows the same expression as Eq.~\eqref{as_VQ}, but with $x_{10}$ replaced by the physical size of the daughter dipole created by the evolution step of interest. 

With the 8 free parameters specified above, the small-$x$ helicity global analysis is performed repeatedly for 10 different values of $C^2$ ranging from 0.05 to 2, which contains the trivial choice of $C^2=1$ and the choice $C^2=e^{-2\gamma_E}\approx 0.3$ with $\gamma_E$ being the Euler-Mascheroni constant. The latter is the choice that connects coordinate- and momentum-space running coupling prescriptions~\cite{Kovchegov:2006vj,Lappi:2012vw}. As a result, despite containing a third of the original number of free parameters, the valence-quark initial conditions remain capable of providing a good description of the data. Specifically, $C^2$ ranging from 0.25 to 1.1 yielding equally good fit, with $\chi^2/N_{\text{pts}} = 1.26$ for $C^2=1$ and $\chi^2/N_{\text{pts}} = 1.28$ for $C^2=e^{-2\gamma_E}$~\cite{Adamiak:2025mdy}. Despite a significant increase in $\chi^2$-statistic from $\chi^2/N_{\text{pts}} = 1.03$ achieved using the generalized Born initial conditions~\cite{Adamiak:2023yhz,JAMCollaborationSmall-xAnalysisGroup:2025tfa}, the resulting predictions of asymmetries remain consistent with the results from Ref.~\cite{Adamiak:2023yhz,JAMCollaborationSmall-xAnalysisGroup:2025tfa}, as illustrated in Figs.~S1 and S2 of Ref.~\cite{Adamiak:2025mdy}. 

More importantly, the predictions of quark and gluon hPDFs, together with the $g_1$ structure function, see drastic improvements in the uncertainties. For instance, the total parton spin inside the proton from $x=10^{-5}$ to $x=1$ is~\cite{Adamiak:2025mdy,Adamiak:2023yhz}
\begin{align}
\label{Sq_Sg_result}
    S_q+S_g &= \int\limits_{10^{-5}}^1dx \left[\frac{1}{2}\Delta\Sigma+\Delta G\right] = \begin{cases}
        -0.64 \pm 0.60 &,\;\;\;\text{generalized Born initial conditions} \\
        0.63 \pm 0.10 &,\;\;\;\text{valence quark initial conditions}
    \end{cases} \, ,
\end{align}
which displays a six-fold improvement in uncertainty. This reduction of uncertainty appears sufficient in fixing the signs of gluon hPDF, $\Delta G$, and $g_1$ at small $x$, particularly favoring \emph{positive} $\Delta G$ and \emph{negative} $g_1$~\cite{Adamiak:2025mdy}. This is illustrated by Figs.~1 and 2 of Ref.~\cite{Adamiak:2025mdy}. These signs were undeterministic based on the Born results of Ref.~\cite{Adamiak:2023yhz}.

However, the large difference between the central values in Eq.~\eqref{Sq_Sg_result} without an overlap of uncertainty bands does put into question the physical accuracy of the valence quark model, which utilizes the valence quarks with a gluon emission as the sole components of a proton at moderate $x\sim 0.1$. To examine the possibility in this direction, there is currently a work in progress by the authors to quantify the uncertainty of this valence-quark approximation by including the effects of parton's transverse momentum, together with the additional quark-to-gluon and gluon-to-quark transition amplitudes~\cite{Borden:2024bxa}. Specifically, for the former, since a parton could contain nonzero transverse momentum relative to the proton, its helicity is not necessarily the spin along the direction of proton's momentum. Although the directional difference is expected to be small in the small-$x$ regime, it remains useful to quantify its impact on the initial conditions of polarized dipole amplitude in the valence quark model. This will be performed via Melosh rotation on the proton's wave function~\cite{Schlumpf:1992vq,Brodsky:1994fz}. The result will also provide a baseline in order to assign a prior distribution functions to $a_F$ and $b_F$, which could serve as a more flexible method to add physical input from the valence quark model into the global analysis of helicity at small $x$.





\section{Conclusion and Future Work}
\label{sec:conclusion}

In this article, we summarize the small-$x$ helicity global analysis of Ref.~\cite{Adamiak:2025mdy}, which is a direct application of the helicity-dependent valence quark model developed in Ref.~\cite{Dumitru:2024pcv}. The model employs physical insight from the valence quark picture to dramatically reduce the number of free parameters in the global analysis~\cite{Adamiak:2023yhz,JAMCollaborationSmall-xAnalysisGroup:2025tfa}, resulting in physical predictions with up to six-fold improvement in uncertainties. Although the $\chi^2$-statistic of the fit deteriorates, the overall results remain indicative of a good agreement with the valence quark model and the polarized DIS and SIDIS data~\cite{Adamiak:2025mdy}. 

However, the authors believe that it will be useful to quantify the uncertainty of the valence quark model via the incorporation of Melosh rotation~\cite{Schlumpf:1992vq,Brodsky:1994fz}, together with the inclusion of quark-to-gluon and gluon-to-quark transition amplitudes~\cite{Borden:2024bxa}. Ultimately, this will lead to a new version of global analysis that can also include polarized jet production in proton-proton collision~\cite{JAMCollaborationSmall-xAnalysisGroup:2025tfa}, whose spin asymmetry contains the same polarized dipole amplitudes at small $x$.

In the longer term, the valence quark model can be employed to calculate the initial conditions for fits of other functions whose small-$x$ evolution is available, including the TMDs~\cite{Kovchegov:2022kyy,Santiago:2023rfl,Adamiak:2024khm,Kovchegov:2025gcg}, orbital angular momenta~\cite{Kovchegov:2023yzd,Becker:2026xcp}, gravitational form factors~\cite{Tong:2022zax,Hagiwara:2024wqz} and more~\cite{Kovchegov:2025yyl,Mantysaari:2025mht,Mantysaari:2026zte}.

\section*{Acknowledgments}

The authors would like to thank the organizers for the opportunity to present the work. 
YT is supported by the National Research Council of Thailand (NRCT) via the project number 220677 (contract number N42A690266), by the National Science, Research and Innovation Fund (NSRF) via the Program Management Unit for Human Resources \& Institutional Development, Research and Innovation (grant number B39G680009), and by grants for development of new faculty members, Ratchadaphiseksomphot Fund, Chulalongkorn University. DA is grateful for support from the Wu-Ki Tung Endowed Chair in particle physics.
HM is supported by the Research Council of Finland, the Centre of Excellence in Quark Matter and projects 338263 and 359902, and by the European Research Council (ERC, grant agreements No. ERC-2023-101123801 GlueSatLight and No. ERC-2018-ADG835105 YoctoLHC). 
The content of this article does not reflect the official opinion of the European Union and responsibility for the information and views expressed therein lies entirely with the authors.



\bibliographystyle{JHEP}

\begin{thebibliography}{10}

\bibitem{Adamiak:2025mdy}
D.~Adamiak, H.~M{\"a}ntysaari and Y.~Tawabutr, \emph{{Proton spin from small-x
  with constraints from the valence quark model}},
  \href{https://doi.org/10.1016/j.physletb.2025.139911}{\emph{Phys. Lett. B}
  {\bfseries 870} (2025) 139911}
  [\href{https://arxiv.org/abs/2502.16604}{{\ttfamily 2502.16604}}].

\bibitem{Dumitru:2024pcv}
A.~Dumitru, H.~M{\"a}ntysaari and Y.~Tawabutr, \emph{{Polarized dipole
  scattering amplitudes meet the valence quark model}},
  \href{https://doi.org/10.1103/PhysRevD.110.054030}{\emph{Phys. Rev. D}
  {\bfseries 110} (2024) 054030}
  [\href{https://arxiv.org/abs/2407.08893}{{\ttfamily 2407.08893}}].

\bibitem{Cougoulic:2022gbk}
F.~Cougoulic, Y.V.~Kovchegov, A.~Tarasov and Y.~Tawabutr, \emph{{Quark and
  gluon helicity evolution at small x: revised and updated}},
  \href{https://doi.org/10.1007/JHEP07(2022)095}{\emph{JHEP} {\bfseries 07}
  (2022) 095} [\href{https://arxiv.org/abs/2204.11898}{{\ttfamily
  2204.11898}}].

\bibitem{Adamiak:2023yhz}
{\scshape Jefferson Lab Angular Momentum (JAM)} collaboration, \emph{{Global
  analysis of polarized DIS and SIDIS data with improved small-x helicity
  evolution}}, \href{https://doi.org/10.1103/PhysRevD.108.114007}{\emph{Phys.
  Rev. D} {\bfseries 108} (2023) 114007}
  [\href{https://arxiv.org/abs/2308.07461}{{\ttfamily 2308.07461}}].

\bibitem{Ashman:1987hv}
{\scshape European Muon} collaboration, \emph{{A Measurement of the Spin
  Asymmetry and Determination of the Structure Function g(1) in Deep Inelastic
  Muon-Proton Scattering}},
  \href{https://doi.org/10.1016/0370-2693(88)91523-7}{\emph{Phys. Lett.}
  {\bfseries B206} (1988) 364}.

\bibitem{Ashman:1989ig}
{\scshape European Muon} collaboration, \emph{{An Investigation of the Spin
  Structure of the Proton in Deep Inelastic Scattering of Polarized Muons on
  Polarized Protons}},
  \href{https://doi.org/10.1016/0550-3213(89)90089-8}{\emph{Nucl. Phys.}
  {\bfseries B328} (1989) 1}.

\bibitem{Jaffe:1989jz}
R.L.~Jaffe and A.~Manohar, \emph{{The G(1) Problem: Fact and Fantasy on the
  Spin of the Proton}},
  \href{https://doi.org/10.1016/0550-3213(90)90506-9}{\emph{Nucl. Phys.}
  {\bfseries B337} (1990) 509}.

\bibitem{Bashinsky:1998if}
S.~Bashinsky and R.L.~Jaffe, \emph{{Quark and gluon orbital angular momentum
  and spin in hard processes}},
  \href{https://doi.org/10.1016/S0550-3213(98)00559-8}{\emph{Nucl. Phys.}
  {\bfseries B536} (1998) 303}
  [\href{https://arxiv.org/abs/hep-ph/9804397}{{\ttfamily hep-ph/9804397}}].

\bibitem{Leader:2013jra}
E.~Leader and C.~Lorcé, \emph{{The angular momentum controversy: What's it all
  about and does it matter?}},
  \href{https://doi.org/10.1016/j.physrep.2014.02.010}{\emph{Phys. Rept.}
  {\bfseries 541} (2014) 163}
  [\href{https://arxiv.org/abs/1309.4235}{{\ttfamily 1309.4235}}].

\bibitem{Gelis:2010nm}
F.~Gelis, E.~Iancu, J.~Jalilian-Marian and R.~Venugopalan, \emph{{The Color
  Glass Condensate}},
  \href{https://doi.org/10.1146/annurev.nucl.010909.083629}{\emph{Ann.Rev.Nucl.Part.Sci.}
  {\bfseries 60} (2010) 463} [\href{https://arxiv.org/abs/1002.0333}{{\ttfamily
  1002.0333}}].

\bibitem{Kovchegov:2012mbw}
Y.V.~Kovchegov and E.~Levin, \emph{{Quantum chromodynamics at high energy}},
  vol.~33, Cambridge University Press (2012).

\bibitem{Balitsky:2008zz}
I.~Balitsky and G.A.~Chirilli, \emph{{Next-to-leading order evolution of color
  dipoles}}, \href{https://doi.org/10.1103/PhysRevD.77.014019}{\emph{Phys.
  Rev.} {\bfseries D77} (2008) 014019}
  [\href{https://arxiv.org/abs/0710.4330}{{\ttfamily 0710.4330}}].

\bibitem{Adamiak:2024khm}
D.~Adamiak, M.G.~Santiago and Y.~Tawabutr, \emph{{Small-x asymptotics of the
  leading-twist flavor-singlet quark TMDs}},
  \href{https://doi.org/10.1103/fyld-m5g1}{\emph{Phys. Rev. D} {\bfseries 113}
  (2026) 014023} [\href{https://arxiv.org/abs/2412.14154}{{\ttfamily
  2412.14154}}].

\bibitem{Albacete:2010sy}
J.L.~Albacete, N.~Armesto, J.G.~Milhano, P.~Quiroga-Arias and C.A.~Salgado,
  \emph{{AAMQS: A non-linear QCD analysis of new HERA data at small-x including
  heavy quarks}},
  \href{https://doi.org/10.1140/epjc/s10052-011-1705-3}{\emph{Eur. Phys. J.}
  {\bfseries C71} (2011) 1705}
  [\href{https://arxiv.org/abs/1012.4408}{{\ttfamily 1012.4408}}].

\bibitem{Beuf:2020dxl}
G.~Beuf, H.~H\"anninen, T.~Lappi and H.~M\"antysaari, \emph{{Color Glass
  Condensate at next-to-leading order meets HERA data}},
  \href{https://doi.org/10.1103/PhysRevD.102.074028}{\emph{Phys. Rev. D}
  {\bfseries 102} (2020) 074028}
  [\href{https://arxiv.org/abs/2007.01645}{{\ttfamily 2007.01645}}].

\bibitem{Adamiak:2021ppq}
{\scshape Jefferson Lab Angular Momentum} collaboration, \emph{{First analysis
  of world polarized DIS data with small-x helicity evolution}},
  \href{https://doi.org/10.1103/PhysRevD.104.L031501}{\emph{Phys. Rev. D}
  {\bfseries 104} (2021) L031501}
  [\href{https://arxiv.org/abs/2102.06159}{{\ttfamily 2102.06159}}].

\bibitem{JAMCollaborationSmall-xAnalysisGroup:2025tfa}
{\scshape JAM Collaboration (Small-x Analysis Group)} collaboration,
  \emph{{First study of polarized proton-proton scattering with small-x
  helicity evolution}}, \href{https://doi.org/10.1103/9gnx-ycs4}{\emph{Phys.
  Rev. D} {\bfseries 112} (2025) 094032}
  [\href{https://arxiv.org/abs/2503.21006}{{\ttfamily 2503.21006}}].

\bibitem{Kovchegov:2015pbl}
Y.V.~Kovchegov, D.~Pitonyak and M.D.~Sievert, \emph{{Helicity Evolution at
  Small-x}}, \href{https://doi.org/10.1007/JHEP01(2016)072}{\emph{JHEP}
  {\bfseries 01} (2016) 072}
  [\href{https://arxiv.org/abs/1511.06737}{{\ttfamily 1511.06737}}].

\bibitem{Kovchegov:2016zex}
Y.V.~Kovchegov, D.~Pitonyak and M.D.~Sievert, \emph{{Helicity Evolution at
  Small $x$: Flavor Singlet and Non-Singlet Observables}},
  \href{https://doi.org/10.1103/PhysRevD.95.014033}{\emph{Phys. Rev.}
  {\bfseries D95} (2017) 014033}
  [\href{https://arxiv.org/abs/1610.06197}{{\ttfamily 1610.06197}}].

\bibitem{Kovchegov:2018znm}
Y.V.~Kovchegov and M.D.~Sievert, \emph{{Small-$x$ Helicity Evolution: an
  Operator Treatment}},
  \href{https://doi.org/10.1103/PhysRevD.99.054032}{\emph{Phys. Rev.}
  {\bfseries D99} (2019) 054032}
  [\href{https://arxiv.org/abs/1808.09010}{{\ttfamily 1808.09010}}].

\bibitem{Kovchegov:2021lvz}
Y.V.~Kovchegov, A.~Tarasov and Y.~Tawabutr, \emph{{Helicity evolution at small
  x: the single-logarithmic contribution}},
  \href{https://doi.org/10.1007/JHEP03(2022)184}{\emph{JHEP} {\bfseries 03}
  (2022) 184} [\href{https://arxiv.org/abs/2104.11765}{{\ttfamily
  2104.11765}}].

\bibitem{Accardi:2012qut}
A.~Accardi et~al., \emph{{Electron Ion Collider: The Next QCD Frontier}},
  \href{https://doi.org/10.1140/epja/i2016-16268-9}{\emph{Eur. Phys. J.}
  {\bfseries A52} (2016) 268}
  [\href{https://arxiv.org/abs/1212.1701}{{\ttfamily 1212.1701}}].

\bibitem{AbdulKhalek:2021gbh}
R.~Abdul~Khalek et~al., \emph{{Science Requirements and Detector Concepts for
  the Electron-Ion Collider}: {EIC Yellow Report}},
  \href{https://doi.org/10.1016/j.nuclphysa.2022.122447}{\emph{Nucl. Phys. A}
  {\bfseries 1026} (2022) 122447}
  [\href{https://arxiv.org/abs/2103.05419}{{\ttfamily 2103.05419}}].

\bibitem{Abir:2023fpo}
R.~Abir et~al., \emph{{The case for an EIC Theory Alliance: Theoretical
  Challenges of the EIC}},  \href{https://arxiv.org/abs/2305.14572}{{\ttfamily
  2305.14572}}.

\bibitem{Dumitru:2018vpr}
A.~Dumitru, G.A.~Miller and R.~Venugopalan, \emph{{Extracting many-body color
  charge correlators in the proton from exclusive DIS at large Bjorken x}},
  \href{https://doi.org/10.1103/PhysRevD.98.094004}{\emph{Phys. Rev. D}
  {\bfseries 98} (2018) 094004}
  [\href{https://arxiv.org/abs/1808.02501}{{\ttfamily 1808.02501}}].

\bibitem{Dumitru:2020fdh}
A.~Dumitru, V.~Skokov and T.~Stebel, \emph{{Subfemtometer scale color charge
  correlations in the proton}},
  \href{https://doi.org/10.1103/PhysRevD.101.054004}{\emph{Phys. Rev. D}
  {\bfseries 101} (2020) 054004}
  [\href{https://arxiv.org/abs/2001.04516}{{\ttfamily 2001.04516}}].

\bibitem{Dumitru:2020gla}
A.~Dumitru and R.~Paatelainen, \emph{{Sub-femtometer scale color charge
  fluctuations in a proton made of three quarks and a gluon}},
  \href{https://doi.org/10.1103/PhysRevD.103.034026}{\emph{Phys. Rev. D}
  {\bfseries 103} (2021) 034026}
  [\href{https://arxiv.org/abs/2010.11245}{{\ttfamily 2010.11245}}].

\bibitem{Borden:2024bxa}
J.~Borden, Y.V.~Kovchegov and M.~Li, \emph{{Helicity evolution at small x:
  quark to gluon and gluon to quark transition operators}},
  \href{https://doi.org/10.1007/JHEP09(2024)037}{\emph{JHEP} {\bfseries 09}
  (2024) 037} [\href{https://arxiv.org/abs/2406.11647}{{\ttfamily
  2406.11647}}].

\bibitem{Veneziano:1976wm}
G.~Veneziano, \emph{{Some Aspects of a Unified Approach to Gauge, Dual and
  Gribov Theories}},
  \href{https://doi.org/10.1016/0550-3213(76)90412-0}{\emph{Nucl. Phys. B}
  {\bfseries 117} (1976) 519}.

\bibitem{Schlumpf:1992vq}
F.~Schlumpf, \emph{{Relativistic constituent quark model of electroweak
  properties of baryons}},
  \href{https://doi.org/10.1103/PhysRevD.47.4114}{\emph{Phys. Rev. D}
  {\bfseries 47} (1993) 4114}
  [\href{https://arxiv.org/abs/hep-ph/9212250}{{\ttfamily hep-ph/9212250}}].

\bibitem{Brodsky:1994fz}
S.J.~Brodsky and F.~Schlumpf, \emph{{Wave function independent relations
  between the nucleon axial coupling g(A) and the nucleon magnetic moments}},
  \href{https://doi.org/10.1016/0370-2693(94)90525-8}{\emph{Phys. Lett. B}
  {\bfseries 329} (1994) 111}
  [\href{https://arxiv.org/abs/hep-ph/9402214}{{\ttfamily hep-ph/9402214}}].

\bibitem{E142:1996thl}
{\scshape E142} collaboration, \emph{{Deep inelastic scattering of polarized
  electrons by polarized He-3 and the study of the neutron spin structure}},
  \href{https://doi.org/10.1103/PhysRevD.54.6620}{\emph{Phys. Rev. D}
  {\bfseries 54} (1996) 6620}
  [\href{https://arxiv.org/abs/hep-ex/9610007}{{\ttfamily hep-ex/9610007}}].

\bibitem{E143:1998hbs}
{\scshape E143} collaboration, \emph{{Measurements of the proton and deuteron
  spin structure functions $g_1$ and $g_2$}},
  \href{https://doi.org/10.1103/PhysRevD.58.112003}{\emph{Phys. Rev. D}
  {\bfseries 58} (1998) 112003}
  [\href{https://arxiv.org/abs/hep-ph/9802357}{{\ttfamily hep-ph/9802357}}].

\bibitem{E154:1997xfa}
{\scshape E154} collaboration, \emph{{Precision determination of the neutron
  spin structure function $g_1^n$}},
  \href{https://doi.org/10.1103/PhysRevLett.79.26}{\emph{Phys. Rev. Lett.}
  {\bfseries 79} (1997) 26}
  [\href{https://arxiv.org/abs/hep-ex/9705012}{{\ttfamily hep-ex/9705012}}].

\bibitem{E155:1999pwm}
{\scshape E155} collaboration, \emph{{Measurement of the deuteron spin
  structure function $g_1^d(x)$ for $1\,(\mathrm{GeV}/c)^2< Q^2 < 40\,
  (\mathrm{GeV}/c)^2$}},
  \href{https://doi.org/10.1016/S0370-2693(99)00940-5}{\emph{Phys. Lett. B}
  {\bfseries 463} (1999) 339}
  [\href{https://arxiv.org/abs/hep-ex/9904002}{{\ttfamily hep-ex/9904002}}].

\bibitem{E155:2000qdr}
{\scshape E155} collaboration, \emph{{Measurements of the $Q^2$ dependence of
  the proton and neutron spin structure functions $g_1^p$ and $g_1^n$}},
  \href{https://doi.org/10.1016/S0370-2693(00)01014-5}{\emph{Phys. Lett. B}
  {\bfseries 493} (2000) 19}
  [\href{https://arxiv.org/abs/hep-ph/0007248}{{\ttfamily hep-ph/0007248}}].

\bibitem{EuropeanMuon:1989yki}
{\scshape European Muon} collaboration, \emph{{An Investigation of the Spin
  Structure of the Proton in Deep Inelastic Scattering of Polarized Muons on
  Polarized Protons}},
  \href{https://doi.org/10.1016/0550-3213(89)90089-8}{\emph{Nucl. Phys. B}
  {\bfseries 328} (1989) 1}.

\bibitem{SpinMuon:1997yns}
{\scshape Spin Muon} collaboration, \emph{{Polarized quark distributions in the
  nucleon from semiinclusive spin asymmetries}},
  \href{https://doi.org/10.1016/S0370-2693(97)01546-3}{\emph{Phys. Lett. B}
  {\bfseries 420} (1998) 180}
  [\href{https://arxiv.org/abs/hep-ex/9711008}{{\ttfamily hep-ex/9711008}}].

\bibitem{SpinMuon:1998eqa}
{\scshape Spin Muon} collaboration, \emph{{Spin asymmetries $A_1$ and structure
  functions $g_1$ of the proton and the deuteron from polarized high-energy
  muon scattering}},
  \href{https://doi.org/10.1103/PhysRevD.58.112001}{\emph{Phys. Rev. D}
  {\bfseries 58} (1998) 112001}.

\bibitem{SpinMuon:1999udj}
{\scshape Spin Muon} collaboration, \emph{{Spin asymmetries $A_1$ of the proton
  and the deuteron in the low $x$ and low $Q^2$ region from polarized
  high-energy muon scattering}},
  \href{https://doi.org/10.1103/PhysRevD.60.072004}{\emph{Phys. Rev. D}
  {\bfseries 60} (1999) 072004}.

\bibitem{COMPASS:2009kiy}
{\scshape COMPASS} collaboration, \emph{{Flavour Separation of Helicity
  Distributions from Deep Inelastic Muon-Deuteron Scattering}},
  \href{https://doi.org/10.1016/j.physletb.2009.08.065}{\emph{Phys. Lett. B}
  {\bfseries 680} (2009) 217}
  [\href{https://arxiv.org/abs/0905.2828}{{\ttfamily 0905.2828}}].

\bibitem{COMPASS:2010hwr}
{\scshape COMPASS} collaboration, \emph{{Quark helicity distributions from
  longitudinal spin asymmetries in muon-proton and muon-deuteron scattering}},
  \href{https://doi.org/10.1016/j.physletb.2010.08.034}{\emph{Phys. Lett. B}
  {\bfseries 693} (2010) 227}
  [\href{https://arxiv.org/abs/1007.4061}{{\ttfamily 1007.4061}}].

\bibitem{COMPASS:2010wkz}
{\scshape COMPASS} collaboration, \emph{{The Spin-dependent Structure Function
  of the Proton $g_1^p$ and a Test of the Bjorken Sum Rule}},
  \href{https://doi.org/10.1016/j.physletb.2010.05.069}{\emph{Phys. Lett. B}
  {\bfseries 690} (2010) 466}
  [\href{https://arxiv.org/abs/1001.4654}{{\ttfamily 1001.4654}}].

\bibitem{COMPASS:2015mhb}
{\scshape COMPASS} collaboration, \emph{{The spin structure function $g_1^{\rm
  p}$ of the proton and a test of the Bjorken sum rule}},
  \href{https://doi.org/10.1016/j.physletb.2015.11.064}{\emph{Phys. Lett. B}
  {\bfseries 753} (2016) 18}
  [\href{https://arxiv.org/abs/1503.08935}{{\ttfamily 1503.08935}}].

\bibitem{COMPASS:2016jwv}
{\scshape COMPASS} collaboration, \emph{{Final COMPASS results on the deuteron
  spin-dependent structure function $g_1^{\rm d}$ and the Bjorken sum rule}},
  \href{https://doi.org/10.1016/j.physletb.2017.03.018}{\emph{Phys. Lett. B}
  {\bfseries 769} (2017) 34}
  [\href{https://arxiv.org/abs/1612.00620}{{\ttfamily 1612.00620}}].

\bibitem{HERMES:1997hjr}
{\scshape HERMES} collaboration, \emph{{Measurement of the neutron spin
  structure function g1(n) with a polarized He-3 internal target}},
  \href{https://doi.org/10.1016/S0370-2693(97)00611-4}{\emph{Phys. Lett. B}
  {\bfseries 404} (1997) 383}
  [\href{https://arxiv.org/abs/hep-ex/9703005}{{\ttfamily hep-ex/9703005}}].

\bibitem{HERMES:1999uyx}
{\scshape HERMES} collaboration, \emph{{Flavor decomposition of the polarized
  quark distributions in the nucleon from inclusive and semiinclusive deep
  inelastic scattering}},
  \href{https://doi.org/10.1016/S0370-2693(99)00964-8}{\emph{Phys. Lett. B}
  {\bfseries 464} (1999) 123}
  [\href{https://arxiv.org/abs/hep-ex/9906035}{{\ttfamily hep-ex/9906035}}].

\bibitem{HERMES:2004zsh}
{\scshape HERMES} collaboration, \emph{{Quark helicity distributions in the
  nucleon for up, down, and strange quarks from semi-inclusive deep-inelastic
  scattering}}, \href{https://doi.org/10.1103/PhysRevD.71.012003}{\emph{Phys.
  Rev. D} {\bfseries 71} (2005) 012003}
  [\href{https://arxiv.org/abs/hep-ex/0407032}{{\ttfamily hep-ex/0407032}}].

\bibitem{HERMES:2006jyl}
{\scshape HERMES} collaboration, \emph{{Precise determination of the spin
  structure function g(1) of the proton, deuteron and neutron}},
  \href{https://doi.org/10.1103/PhysRevD.75.012007}{\emph{Phys. Rev. D}
  {\bfseries 75} (2007) 012007}
  [\href{https://arxiv.org/abs/hep-ex/0609039}{{\ttfamily hep-ex/0609039}}].

\bibitem{Kovchegov:2006vj}
Y.~Kovchegov and H.~Weigert, \emph{{Triumvirate of Running Couplings in
  Small-$x$ Evolution}}, {\emph{Nucl. Phys. {\bf A}} {\bfseries 784} (2007)
  188} [\href{https://arxiv.org/abs/hep-ph/0609090}{{\ttfamily
  hep-ph/0609090}}].

\bibitem{Lappi:2012vw}
T.~Lappi and H.~M{\"a}ntysaari, \emph{{On the running coupling in the JIMWLK
  equation}}, \href{https://doi.org/10.1140/epjc/s10052-013-2307-z}{\emph{Eur.
  Phys. J. C} {\bfseries 73} (2013) 2307}
  [\href{https://arxiv.org/abs/1212.4825}{{\ttfamily 1212.4825}}].

\bibitem{Kovchegov:2022kyy}
Y.V.~Kovchegov and M.G.~Santiago, \emph{{T-odd leading-twist quark TMDs at
  small x}}, \href{https://doi.org/10.1007/JHEP11(2022)098}{\emph{JHEP}
  {\bfseries 11} (2022) 098}
  [\href{https://arxiv.org/abs/2209.03538}{{\ttfamily 2209.03538}}].

\bibitem{Santiago:2023rfl}
M.G.~Santiago, \emph{{Spin-spin coupling at small x: Worm-gear and pretzelosity
  TMDs}}, \href{https://doi.org/10.1103/PhysRevD.109.034004}{\emph{Phys. Rev.
  D} {\bfseries 109} (2024) 034004}
  [\href{https://arxiv.org/abs/2310.02231}{{\ttfamily 2310.02231}}].

\bibitem{Kovchegov:2025gcg}
Y.V.~Kovchegov and M.~Li, \emph{{Weizs{\"a}cker-Williams gluon helicity
  distribution and inclusive dijet production in longitudinally polarized
  electron-proton collisions}},
  \href{https://doi.org/10.1007/JHEP08(2025)206}{\emph{JHEP} {\bfseries 08}
  (2025) 206} [\href{https://arxiv.org/abs/2504.12979}{{\ttfamily
  2504.12979}}].

\bibitem{Kovchegov:2023yzd}
Y.V.~Kovchegov and B.~Manley, \emph{{Orbital angular momentum at small x
  revisited}}, \href{https://doi.org/10.1007/JHEP02(2024)060}{\emph{JHEP}
  {\bfseries 02} (2024) 060}
  [\href{https://arxiv.org/abs/2310.18404}{{\ttfamily 2310.18404}}].

\bibitem{Becker:2026xcp}
G.Z.~Becker, Y.~Kovchegov, V., M.~Li, B.~Manley and A.~Tarasov, \emph{{Orbital
  angular momentum at small $x$ in the large $N_c\&N_f$ limit}},
  \href{https://arxiv.org/abs/2608.25120}{{\ttfamily 2608.25120}}.

\bibitem{Tong:2022zax}
X.-B.~Tong, J.-P.~Ma and F.~Yuan, \emph{{Perturbative calculations of
  gravitational form factors at large momentum transfer}},
  \href{https://doi.org/10.1007/JHEP10(2022)046}{\emph{JHEP} {\bfseries 10}
  (2022) 046} [\href{https://arxiv.org/abs/2203.13493}{{\ttfamily
  2203.13493}}].

\bibitem{Hagiwara:2024wqz}
Y.~Hagiwara, X.-B.~Tong and B.-W.~Xiao, \emph{{Understanding gravitational form
  factors with the Weizs{\"a}cker-Williams method}},
  \href{https://doi.org/10.1103/PhysRevD.111.L051503}{\emph{Phys. Rev. D}
  {\bfseries 111} (2025) L051503}
  [\href{https://arxiv.org/abs/2401.12840}{{\ttfamily 2401.12840}}].

\bibitem{Kovchegov:2025yyl}
Y.V.~Kovchegov, M.G.~Santiago and H.~Sun, \emph{{Unpolarized GPDs at small x
  and non-zero skewness}},
  \href{https://doi.org/10.1016/j.physletb.2026.140470}{\emph{Phys. Lett. B}
  {\bfseries 877} (2026) 140470}
  [\href{https://arxiv.org/abs/2512.10086}{{\ttfamily 2512.10086}}].

\bibitem{Mantysaari:2025mht}
H.~M{\"a}ntysaari, Y.~Tawabutr and X.-B.~Tong, \emph{{Nucleon energy
  correlators for the odderon}},
  \href{https://doi.org/10.1103/rmzx-tgm2}{\emph{Phys. Rev. D} {\bfseries 112}
  (2025) 114027} [\href{https://arxiv.org/abs/2503.20157}{{\ttfamily
  2503.20157}}].

\bibitem{Mantysaari:2026zte}
H.~M{\"a}ntysaari, Y.~Shi, Y.~Tawabutr and X.-B.~Tong, \emph{{Gluonic nucleon
  energy correlators and fracture functions for Color Glass Condensate}},
  \href{https://arxiv.org/abs/2608.10955}{{\ttfamily 2608.10955}}.

\end{thebibliography}

\providecommand{\href}[2]{#2}\begingroup\raggedright\endgroup



\end{document}